\documentclass{PoS}
\usepackage{mathtools}

\title{Relating LHC event rates to supersymmetric Grand Unified Theories containing $SU(5)$}
\ShortTitle{Relating LHC event rates to supersymmetric Grand Unified Theories containing $SU(5)$}

\author{\speaker{Bj\"orn Herrmann}, Yannick Stoll \\
  LAPTh, Universit\'e Savoie Mont Blanc, CNRS, 9 Chemin de Bellevue, F-74941 Annecy-le-Vieux, France\\
  E-mail: \email{herrmann@lapth.cnrs.fr}}

\author{Sylvain Fichet \\
  ICTP South American Institute for Fundamental Research, Instituto de Fisica Teorica, Sao Paulo State University, Brazil \\
  International Institute of Physics, UFRN, Av.\ Odilon Gomes de Lima, 1722 - Capim Macio - 59078-400 - Natal-RN, Brazil}

\abstract{We elaborate on a recently found $SU(5)$ relation confined to the up-(s)quark flavour space, that remains immune to large quantum corrections up to the TeV scale. We investigate the possibilities opened by this new window on the GUT scale in order to find TeV-scale $SU(5)$ tests realizable at the LHC. We present a variety of tests, which appear as relations among observables involving flavour violation or chirality flips and rely on the techniques of top polarimetry, charm-tagging, or Higgs detection from cascade decays. We discuss the cases of natural Supersymmetry and top-charm Supersymmetry as example cases. We find that $O(10)$ to $O(100)$ events are needed to obtain 50\% of relative precision at 3$\sigma$ significance for all proposed tests.}

\FullConference{The European Physical Society Conference on High Energy Physics -- EPS-HEP2015\\
		22-29 July 2015\\
		Wien, Austria}

\begin{document}

\section{Introduction: $SU(5)$ relations in the up-squark sector} 
\label{Sec:Intro}

A fairly fascinating feature of the Standard Model (SM) of particle physics is that the matter fields fit into a complete representation of the gauge groupe $SU(5)$ \cite{Georgi:1974sy}. More precisely, the fields $\{ Q_i, U_i, E_i \}$ and $\{L_i, D_i \}$ can be embedded into three copies of the $\bf{10}$ and $\bf{\bar{5}}$ representations of $SU(5)$. 
This can be seen as a hint that the SM gauge group arises as the low-energy limit of a Grand Unified Theory (GUT) broken to $SU(5)$ at some intermediate scale (see Refs.\ \cite{Fritzsch:1974nn, GellMann:1976pg, Raby:2011jt, Slansky:1981yr} for reviews), i.e.\ 
\begin{equation}
	G_{\rm GUT} ~\to~ SU(5) ~\to~ SU(3) \times SU(2) \times U(1).
\end{equation}

Unravelling whether or not Nature is $SU(5)$-symmetric constitutes a challenging open problem in particle physics. Many realizations of $SU(5)$-like GUTs are possible, with various consequences at low energy. In the present work, we consider the case of Supersymmetry (SUSY). In this framework, new forms of matter exist which may give new insights on $SU(5)$ symmetry. More precisely, the interactions between matter and Higgs fields are described by the superpotential
\begin{equation}
	W ~=~ \lambda_d^{ij} H_d 10_i \bar{5}_j + y_u^{ij} H_u 10_i 10_j 
	~\xrightarrow{SU(5)~{\rm breaking}}~ y_d^{ij} H_d Q_i D_j + y_{\ell}^{ij} H_d L_i E_j + y_u^{ij} H_u Q_i U_j \,,
\end{equation}
where $i,j$ are generation indices. Moreover, the Lagrangian of the theory contains SUSY-breaking scalar mass and trilinear terms,
\begin{equation}
	{\cal L} ~\supset~ -\tilde{q}^* M_{\tilde{Q}}^2 \tilde{q} - -\tilde{d}^* M_{\tilde{D}}^2 \tilde{d} -\tilde{u}^* M_{\tilde{U}}^2 \tilde{u} + a_d H_d \tilde{q} \tilde{d} + a_{\ell} H_d \tilde{\ell} \tilde{e} + a_u H_u \tilde{q} \tilde{u}.
\end{equation}

The fact that the $10_i 10_j$ term in the superpotential is symmetric enforces a symmetric top Yukawa coupling together with a symmetric up-type trilinear coupling at the unification scale,
\begin{equation}
	y_u ~=~ y_u^t \qquad {\rm and} \qquad a_u ~=~ a_u^t.
\end{equation}
It is this last property which provides a potential way to test the $SU(5)$ hypothesis and is at the centre of our interest in the following discussion.

While the above relations are exact at the GUT scale, they are modified when evolving the parameters to the TeV scale through renormalization group (RG) running. In particular, relations between the down-type and lepton Yukawa and trilinear couplings are highly model-dependent. Such relations can therefore hardly constitute a generic test of the $SU(5)$ GUT hypothesis. 

The situation is different in the up-squark sector. At the one-loop level, the relevant beta functions 
are dominated by symmetric terms, such that we can expect the matrices $y_u$ and $a_u$ to remain symmetric to a good precision at energies reachable at the LHC. For a more detailed discussion see Refs.\ \cite{Fichet:2014vha, Fichet:2015oha}.

\begin{figure}
	\begin{center}
		\includegraphics[width=0.47\textwidth]{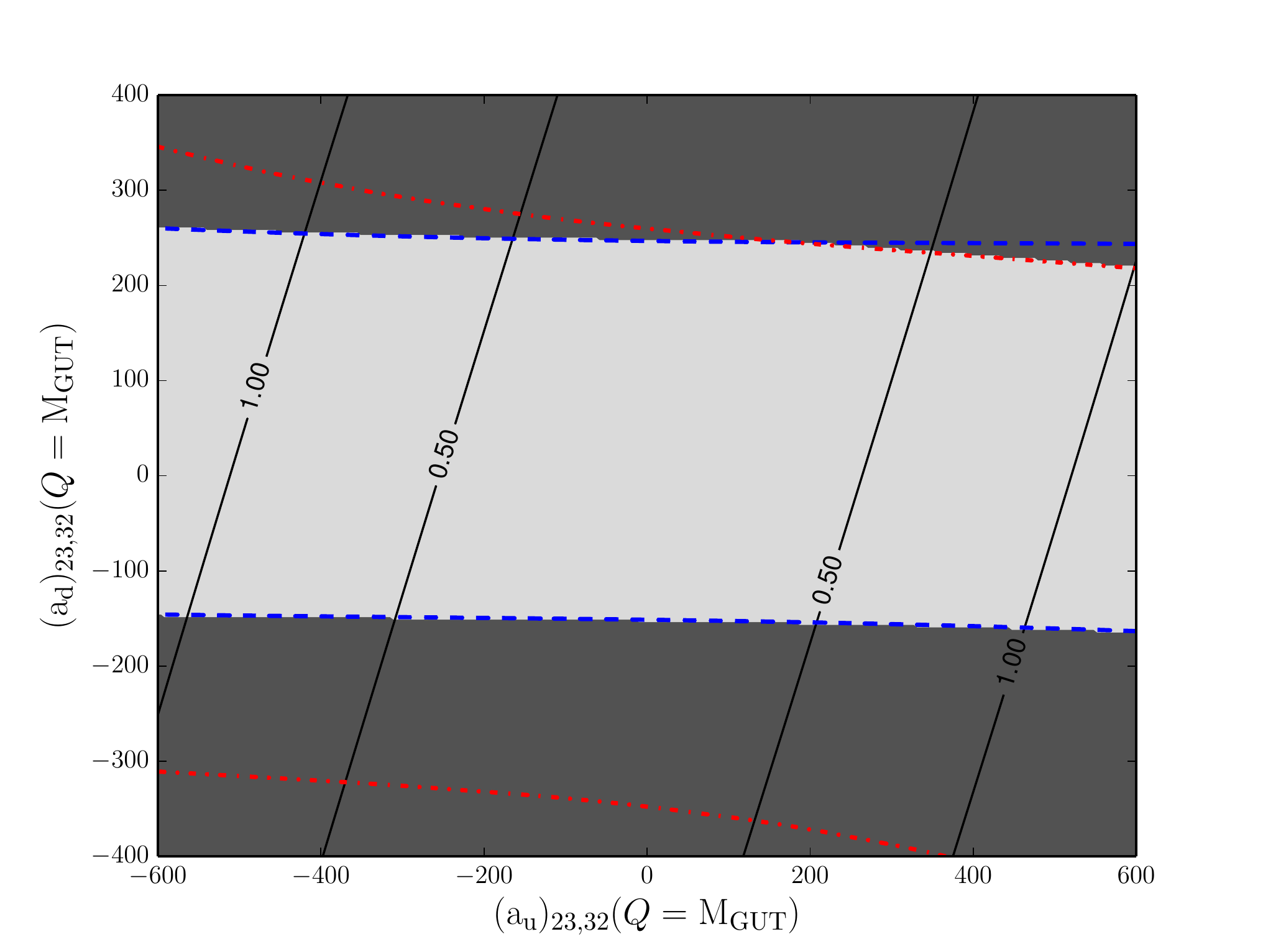} \qquad
		\includegraphics[width=0.47\textwidth]{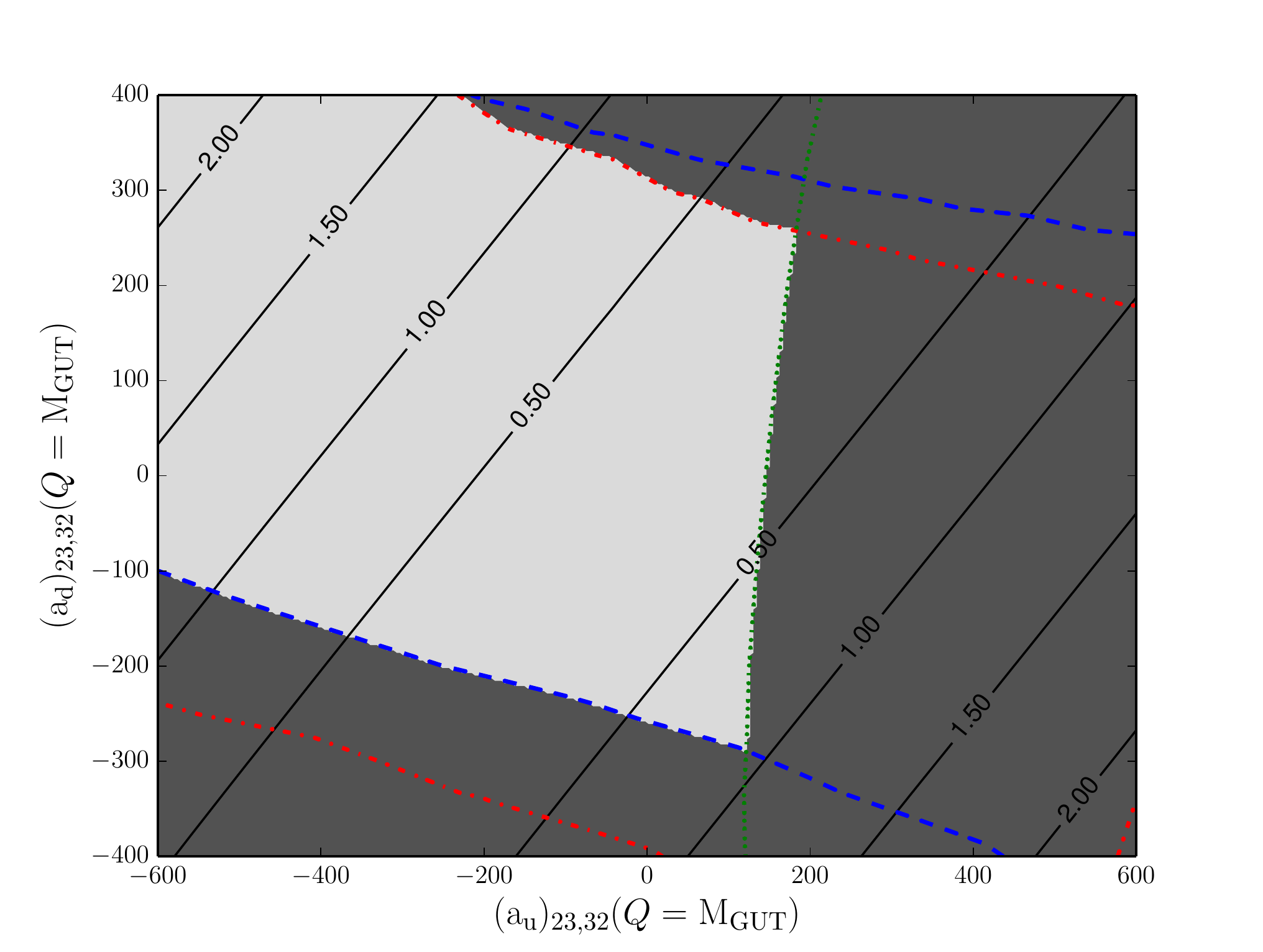}
	\end{center}
	\vspace*{-5mm}
	\caption{The asymmetry $A_{23}$ (black solid line, given in percent) together with the 2$\sigma$ exclusion bands from $\Delta M_{B_s}$ (blue dashed lines) and ${\rm BR}(b\to s\mu\mu)$ (red dash-dotted lines) evaluated for an example scenario with $\tan\beta=10$ (left) and $\tan\beta=40$ (right). The grey area represents the allowed zone taking into account all relevant flavour constraints. For more details on the example scenario and the constraints see Ref.\ \cite{Fichet:2015oha}.}
	\label{Fig:Asymmetry}
\end{figure}

In order to numerically quantify the effect of the RG running, we define the asymmetry at the TeV scale as
\begin{equation}
	A_{ij} ~=~ \frac{ \big| (a_u)_{ij} - (a_u)_{ji} \big| }{{\rm Tr}\big\{ M_{\tilde{u}}^2 \big\}} ,
\end{equation}
where $M_{\tilde{u}}^2$ is the up-squark mass matrix. This quantity does generally not exceed a few percent, as can be seen in Fig.\ \ref{Fig:Asymmetry} for two example cases. 
We can thus summarize our findings by the implication
\begin{equation}
	\big\{ {\rm SUSY~GUT~} G_{\rm GUT} \supset SU(5) \big\} ~\Rightarrow~ \big\{ a_u \approx a_u^t ~{\rm at~the~TeV~scale} \big\} .
\end{equation}
Even stronger, finding a significant difference between $a_u$ and $a_u^t$ at the TeV scale would prove the $SU(5)$ GUT hypothesis wrong. Finally, let us emphasize that the above argumenation remains true for any GUT based on a group $G_{\rm GUT}$, e.g.\ $SO(10)$, containing $SU(5)$ as a subgroup.

\section{Strategies for testing the $SU(5)$ hypothesis at the TeV scale}
\label{Sec:Strategies}

Any strategy that can be set up to test the $SU(5)$ relation discussed above relies on a comparison involving at least two observed up-type squarks. Some of the squarks can be light enough to be produced at the LHC, others may be too heavy such that they appear only virtually in intermediate processes. Depending on the exact pattern of the up-squark mass matrix, two expansions can be used in order to simplify the situation: The mass insertion approximation (MIA) or the effective field theory (EFT) approach. The feasibility of the proposed $SU(5)$ tests depends crucially on the amount of available data, whether they involve real or virtual squarks. 

In many classes of physical models, the squark masses exhibit some hierarchy. The physics of the light squarks can then be captured in an effective Lagrangian, where the heavier states are integrated out. The higher-dimensional operators present in this tree-level Lagrangian will serve as a basis for the $SU(5)$ tests based on both real and virtual squarks. 

When all up-type squarks are heavier than the typical LHC scale, they can only appear off-shell. Potential tests of the $SU(5)$ hypothesis are thus based on virtual squarks only. All squarks can be integrated out and the effective Lagrangian contains the Standard Model plus possibly other light SUSY particles. If the whole SUSY spectrum is heavy, operators of interest are stemming from one-loop diagrams involving at least one Higgs-squark-squark vertex together with fermions on the outgoing legs in order to access information about the flavour structure. We propose $SU(5)$ tests for this situation in Refs.\ \cite{Fichet:2014vha, Fichet:2015oha}. Moreover, tests involving ultraperiphal searches are proposed. 

If the relative mass difference of the squarks is small, the mass insertion approximation (MIA) can be applied. The $SU(5)$ relation of our interest then translates into a relation between the corresponding mass insertions. Note that the MIA is also valid in the case where only a subset of the squark eigenstates is nearly mass degenerate. 

Finally, if the $SU(5)$ tests consists of a simple relation, e.g.\ between event rates at the LHC, a frequentist $p$-value can be used to evaluate the potential of the test. For the tests proposed in Refs.\ \cite{Fichet:2014vha, Fichet:2015oha}, we systematically report the power of the test based on this $p$-value analysis.

\section{Selected results}
\label{Sec:Results}

In this Section we outline the results obtained for the two example cases of natural SUSY and top-charm SUSY. For a detailed discussion and other cases the reader is referred to Refs.\ \cite{Fichet:2014vha, Fichet:2015oha}.

\subsection{Natural supersymmetry}
\label{Sec:NaturalSUSY}

In natural supersymmetry, the mass spectrum features a first and second generation of up-squarks that are considerably heavier than the squarks of the third generation. Thus, the effective theory consists of two squarks, which are mostly stop-like. The effective operators appearing when integrating out the four heavier squarks can potentially induce flavour-changing decays of the stops.

All information concerning the $SU(5)$ relation of our interest, $a_u ~\approx~ a_u^t$, is enclosed in the higher-dimensional operators of the effective Lagrangian for the stops. The associated tests of the $SU(5)$ hypothesis involve both real and virtual squarks. In the following, we assume that both stop mass eigenstates, $\tilde{t}_1$ and $\tilde{t}_2$ are produced at the Large Hadron Collider. The total production cross-section of stop pairs are, at next-to-leading order, $\sigma_{\tilde{t}\tilde{t}} \sim 90$, 8.5, 0.7, and 0.08 fb for stop masses $m_{\tilde{t}} \sim 700$, 1000, 1400, and 1800 GeV, respectively \cite{SUSYXSec}.

As it is expected from a theory with gaugino mass unification at some high scale, we assume the lightest neutralino to be mostly bino-like, $\tilde{\chi}^0_1 \sim \tilde{B}$, while the second-lightest one is mostly wino-like, $\tilde{\chi}^0_2 \sim \tilde{W}$.

If the stops are heavier than the wino, i.e.\ $m_{\tilde{t}} > m_{\tilde{W}} > m_{\tilde{B}}$, they decay directly either into the lightest or into the second-lightest neutralino. In order to build a test of the $SU(5)$ hypothesis, we are interested in the flavour-changing decays
\begin{equation}
	\tilde{t} ~\to~ \tilde{W} ~ u/c ~\to~ \tilde{B} ~ Z/h ~ u/c \quad \mathrm{and} \quad
	\tilde{t} ~\to~ \tilde{B} ~ u/c .
\end{equation}
We assume that only the event rates $N_Y$ and $N_L$ of the stop decays into binos and winos are experimentally accessible, while all other information related to the stops, in particular their masses, is unknown. Moreover, we assume that a certain fraction $N_Y^c$ and $N_L^c$ of these events can be charm-tagged. The remaining events $N_Y^{\not{c}} = N_Y - N_Y^c$ and $N_L^{\not{c}} = N_L - N_L^c$ then contain both up-quark events and miss-tagged charm jets. 

Assuming the same charm-tagging efficiency for $N_Y$ and $N_L$ leads to the relation
\begin{equation}
	\frac{N_Y^c}{N_L^c} ~=~ \frac{N_Y^{\not{c}}}{N_L^{\not{c}}} .
\end{equation}
This relation is easily generalised in case of different charm-tagging efficiencies. The relatively large QCD error on the underlying production cross-sections roughly cancels out in the ratios of event rates. Note that no information on the stop mixing angle nor the stop masses is necessary to carry out the test.

Testing this relation with 50\%, 20\%, or 10\% accuracy at 3$\sigma$ confidence level requires $N_Y \sim N_L \sim 110$, 675, or 2700 events, respectively. For comparison, assuming stop flavour-violating branching ratios of 0.05 and an integrated luminosity of 300 fb$^{-1}$, we can expect about 1340, 130, and 11 events for $m_{\tilde{t}} = 700$, 1000, and 1400 GeV, respectively. 

If the stops are lighter than the wino, i.e.\ $m_{\tilde{W}} > m_{\tilde{t}} > m_{\tilde{B}}$, they only decay into the bino according to 
\begin{equation}
	\tilde{t} \to t \tilde{B}.
\end{equation}
Performing top polarimetry on a decaying stop pair then gives potentially access to the stop mixing angle. The same kind of procedure also provides an $SU(5)$ test. Let us assume that the spin of the tops is analyzed through distributions of the form $(1+\kappa P_t z)$ with $z \in [-1;1]$. The decays of the stops $t_a$ and $t_b$ leading to the event rates $N_a$ and $N_b$ are then splitted over the domains $D_- = [-1; 0]$ and $D_+ = [0; 1]$ such that 
\begin{equation}
	N_a ~=~ N_{a+} + N_{a-} \qquad {\rm and} \qquad N_b ~=~ N_{b+} + N_{b-} .
\end{equation}
These event rates satisfy a non-trivial relation (see Ref.\ \cite{Fichet:2015oha} for details) if the $SU(5)$ hypothesis is verified. 

With a spin analyser of efficiency $\kappa = 0.5$ and $N_a = 20$, $N_b \gtrsim 137$ events are needed to probe the relation at 3$\sigma$ confidence level. Testing the relation with 50\% (20\%) presision requires $N_b \approx 589$ (7560) events. For comparison, for an integrated luminosity of 300 fb$^{-1}$, we expect about 26700 (213) events for stop masses of $m_{\tilde{t}} = 700$ (1400) GeV.

\subsection{Top-charm supersymmetry}
\label{Sec:TopCharmSUSY}

The scenario of top-charm supersymmetry features a heavy first generation of up-type squarks, while -- in contrast to the previous case -- both the stops and charm states are accessible at the LHC. We first consider the case where all stop and scharm masses are nearly degenerate. The mass insertion approximation is then valid for the sector of stops and scharms, and these states should be produced in equally abundant way at the LHC. The off-diagonal elements of the trilinear up-type matrix are identified with the mass insertions $(\delta_u^{LR})_{ij}$ and the $SU(5)$ hypothesis implies
\begin{equation}
	(\delta_u^{LR})_{23} ~\approx~ (\delta_u^{LR})_{32} .
\end{equation}
These mass insertions are related to the squark-Higgs coupling, such that one may use the Higgs as a probe of the squark mass eigenstates. The LHC processes of interest are therefore stop and scharm production, followed by flavour-violating decays into a squark and a Higgs in one of the decay chains. We further assume that the squarks decay into the bino. Such processes can be identified requiring a single top, a hard jet, a Higgs, and large missing energy. This test has to rely on the distinction between the chiralities, which is possible experimentally only for the top quark. Provided that the cascade decay with $\tilde{c}_L \to h^0 \tilde{t}_R$ can be isolated, top polarimetry then readily provides a $SU(5)$ since
\begin{equation}
	{\rm BR}( \tilde{c}_L \to h \tilde{t}_R ) ~\propto~ |(\delta_u^{LR})_{23}|^2 \qquad {\rm and} \qquad
	{\rm BR}( \tilde{c}_R \to h \tilde{t}_L ) ~\propto~ |(\delta_u^{LR})_{32}|^2 .
\end{equation}

In the case of large stop mixing, keeping the scharm masses are still nearly degenerate, the MIA applies to the scharm sector only, while the stops have to be rotated in their mass eigenbasis. Assuming the mass hierarchy $m_{\tilde{t}_2} > m_{\tilde{c}} > m_{\tilde{t}_1}$, the SUSY cascade decays are rather different from the situation discussed above. Flavour-changing scharm decays going through $\tilde{t}_2$ are now suppressed such that top polarimetry is not useful any more. The decays
\begin{equation}
	\tilde{t}_2 ~\to~ h^0 \tilde{c} \qquad {\rm and} \qquad \tilde{c} ~\to~ h^0 \tilde{t}_1
\end{equation}
are now open. We require again a single top, a hard jet, a Higgs, and large missing energy from both sides of the decay chains. The $SU(5)$ test can then be built from the numbers $N_{hj}$ and $N_{ht}$ of events corresponding to the two decay modes indicated above. Assuming $N_{hj} \ll N_{ht}$ and $\theta_{\tilde{t}} = 0.4$, testing the relation with 50\% (10\%) precision at 3$\sigma$ confidence level requires $N_{hj} \gtrsim 19$ (464) events, respectively. Roughly twice less events are needed for $\theta_{\tilde{t}} = 0$. For comparison, assuming flavour-violating branching ratios of 0.05 and an integrated luminosity of 300 fb$^{-1}$, we expect about 1340 (11) events for squark masses of 700 (1400) GeV.

\section{Conclusion and outlook}
\label{Sec:Conclusion}

We elaborate on the relation $a_u = a_u^t$ indicating that the up-squark trilinear coupling matrix is symmetric at the GUT scale if one assumes an $SU(5)$-like Grand Unification. This relation is found to survive to good approximation through the MSSM renormalization group evolution, such that it is spoiled by $O(1\%)$ of relative error at the TeV scale, where squarks would be observed. This relation can thus be taken as window to GUT physics. All $SU(5)$ tests of this property have to rely on either flavour violation or chirality flip in the up-type sector. We propose several such tests for the LHC.

In many cases, the tests we find consist in determining whether a relation among certain observables (e.g.\ event rates) is satisfied or not. In order to quantify the feasibility of the test, we introduce a systematic procedure relying on a frequentist $p$-value. The associated expected precision tells, for a given amount of data, up to which magnitude a violation of the $SU(5)$ relation can be assessed within a given statistical significance.

The $SU(5)$ tests proposed in the above publications are summarized in Table \ref{Tab:Summary}. The typical amount of events needed to reach an expected precision of 50\% at 3$\sigma$ is also shown for each test. The number of needed events ranges roughly from 10 to 100. In cases where no simple relation between certain observables can be defined, a more global hypothesis testing has to be performed. This is subject to ongoing work \cite{MCMCPaper}. 

\begin{table}[h]
\centering
\begin{tabular}{|c|c|c|c|c|c|} 
\hline 
  & Heavy & \multicolumn{2}{|c|}{Natural SUSY} & \multicolumn{2}{|c|}{Top-charm SUSY} \\ 
 & SUSY & $m_{\tilde t_{1,2}} \!>\! m_{\tilde B, \tilde W}$  & 
$m_{\tilde W} \!>\! m_{\tilde t_{1,2}} \!>\! m_{\tilde B}$ & 
 $m_{\tilde t_{L,R}} \!\sim\! m_{\tilde c_{L,R}}$ &  $m_{\tilde t_{2}} \!>\! m_{\tilde c_{L,R}} \!>\! m_{\tilde t_{1}}$
  \\ 
  \hline
  Squarks involved & virtual & \multicolumn{2}{|c|}{virtual/real} & \multicolumn{2}{|c|}{real} \\
\hline 
Top polarimetry & yes & no & yes & yes & no \\ 
\hline 
Charm-tagging & no & yes & no & no & no \\ 
\hline 
Higgs detection & no & no & no & yes & yes \\ 
\hline 
$\theta_t$-dependence & no & no & yes & no & yes \\ 
\hline 
$P_3=50\%$ & $144$ & $72$ & $108$ & $144$ & $10$ \\
\hline
\end{tabular} 
\caption{Summary of the $SU(5)$-tests appearing in the various SUSY scenarios considered.
The last line shows the typical number of events needed to reach a $50\%$ precision at 3$\sigma$.}
 \label{Tab:Summary}
\end{table}


\end{document}